\documentclass[doublecol]{epl2}

\usepackage{amsmath,amssymb}

\usepackage{cite}

\newcommand{\ii}{\mathrm{i}}

\DeclareMathOperator{\tr}{tr}

\title{Optomechanical multistability in the quantum regime}

\author{C. Schulz \and A. Alvermann\thanks{E-mail: \email{alvermann@physik.uni-greifswald.de}} \and L. Bakemeier \and H. Fehske}
\shortauthor{C. Schulz \etal}

\institute{                    
 Institute of Physics, Ernst-Moritz-Arndt-University, 17487 Greifswald, Germany
}

\abstract{
Classical optomechanical systems feature self-sustained oscillations, where multiple periodic orbits at different amplitudes coexist.
We study how this multistability is realized in the quantum regime, where new dynamical patterns appear because quantum trajectories can move between different classical orbits.
We explain the resulting quantum dynamics from the phase space point of view,
and provide a quantitative description in terms of autocorrelation functions.
In this way we can identify clear dynamical signatures of the crossover from classical to quantum mechanics in experimentally accessible quantities.
Finally, we discuss a possible interpretation of our results in the sense that quantum mechanics protects optomechanical systems against the chaotic dynamics realized in the classical limit.
}

\begin{document}

\maketitle

\section{Introduction}

The interaction of light with mechanical objects~\cite{KV08,MG09} enjoys continued interest due to the  successful construction and manipulation of optomechanical devices
over a wide range of system sizes and parameter combinations 
(see the recent reviews~\cite{M13,AKM13} and references cited therein).
With these devices both classical non-linear dynamics such as
self-sustained oscillations~\cite{MHG06,LKM08,KRCSV05,RKCV05}
 and chaos~\cite{CRYKV05,CCV07,BAF14_PRL}
as well as quantum mechanical mechanisms such as
cooling into the groundstate~\cite{C_etal11,T_etal11} and quantum non-demolition measurements~\cite{HRNSCS10,VHCA13,SWLWSMCS14} can be studied in a unified experimental setup.

This raises the question whether it might be possible to detect
the crossover from classical to quantum mechanics directly in the dynamical behaviour of optomechanical systems.
In a previous paper~\cite{BAF14_PRL} we observed that the classical dynamical patterns,
which are characterized by the multistability of self-sustained oscillations, change in a characteristic way if one moves into the quantum regime. Previously stable orbits become unstable, the system oscillates at a new amplitude,
and especially the classical chaotic dynamics is almost immediately replaced by simple periodic oscillations.
In this paper we explain this behaviour from the point of view of classical and quantum phase space dynamics.
Most importantly, we will show that the dynamical patterns do not change at random but that clearly identifiable and new signatures can be observed.

The prototypical optomechanical system is a vibrating cantilever subject to the radiation pressure of a cavity photon field, for which the Hamilton operator reads~\cite{Law95,M13,AKM13}
\begin{equation}
\begin{split}
\tfrac1\hbar H = \left[ \Omega_\text{cav} - \Omega_\text{las}  + g_\text{rad}( b^\dagger + b) \right] a^\dagger a  \\ + \Omega b^\dagger b + \alpha_\text{las} (a^\dagger + a) \;,
\end{split}
\label{eq:hamiltonian}
\end{equation}
where $b^{(\dagger)}$ and $a^{(\dagger)}$ are bosonic operators for the vibrational mode of the cantilever (frequency $\Omega$)  and for the cavity photon field ($\Omega_\text{cav}$), respectively.
This Hamilton operator applies to any generic optomechanical system, but we adopt the cavity-cantilever terminology throughout this paper.
For our theoretical analysis we use the quantum optical master equation~\cite{Carm99}
\begin{equation}
\partial_t \rho = - \frac\ii\hbar [H,\rho] + \Gamma \mathcal D [b,\rho]
                          + \kappa \mathcal D [a,\rho]
\label{eq:master}
\end{equation}
for the cantilever-cavity density matrix $\rho(t)$,
with the dissipative terms 
\begin{equation}
 \mathcal D [L,\rho] = L \rho L^\dagger - \dfrac{1}{2} (L^\dagger L \rho + \rho L^\dagger L)
\label{eq:dissipator}
\end{equation}
that account for cantilever damping ($\propto \Gamma$) and radiative losses ($\propto \kappa$).
Note that the above Hamilton operator is given in a frame that rotates with the frequency $\Omega_\text{las}$ of the external pump laser such that only  the cavity-laser detuning $\Omega_\text{cav} - \Omega_\text{las}$ appears, and that we assume zero temperature in the master equation.

\begin{figure}
\includegraphics[width=0.47\linewidth]{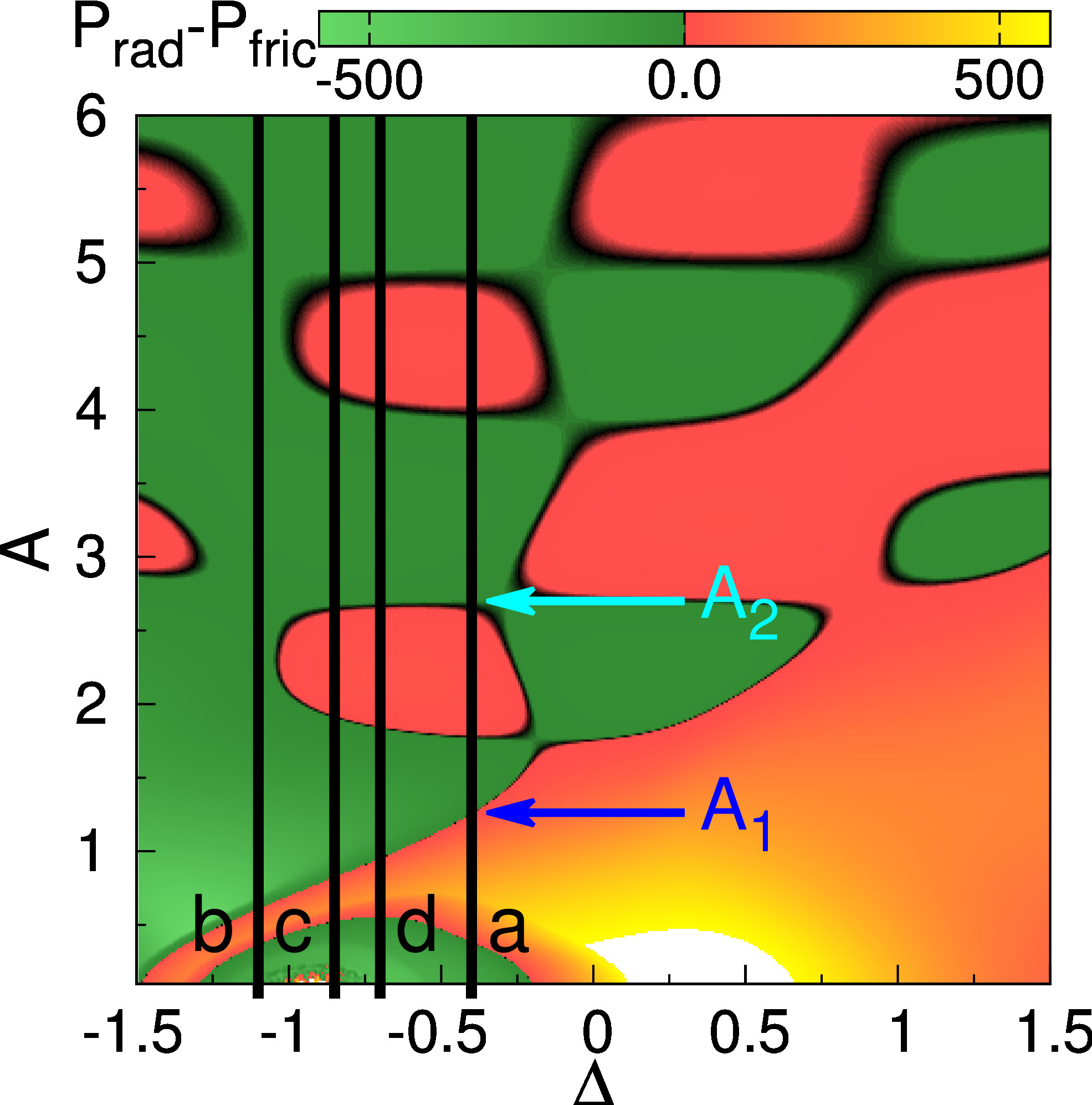}
\hfill
\includegraphics[width=0.47\linewidth]{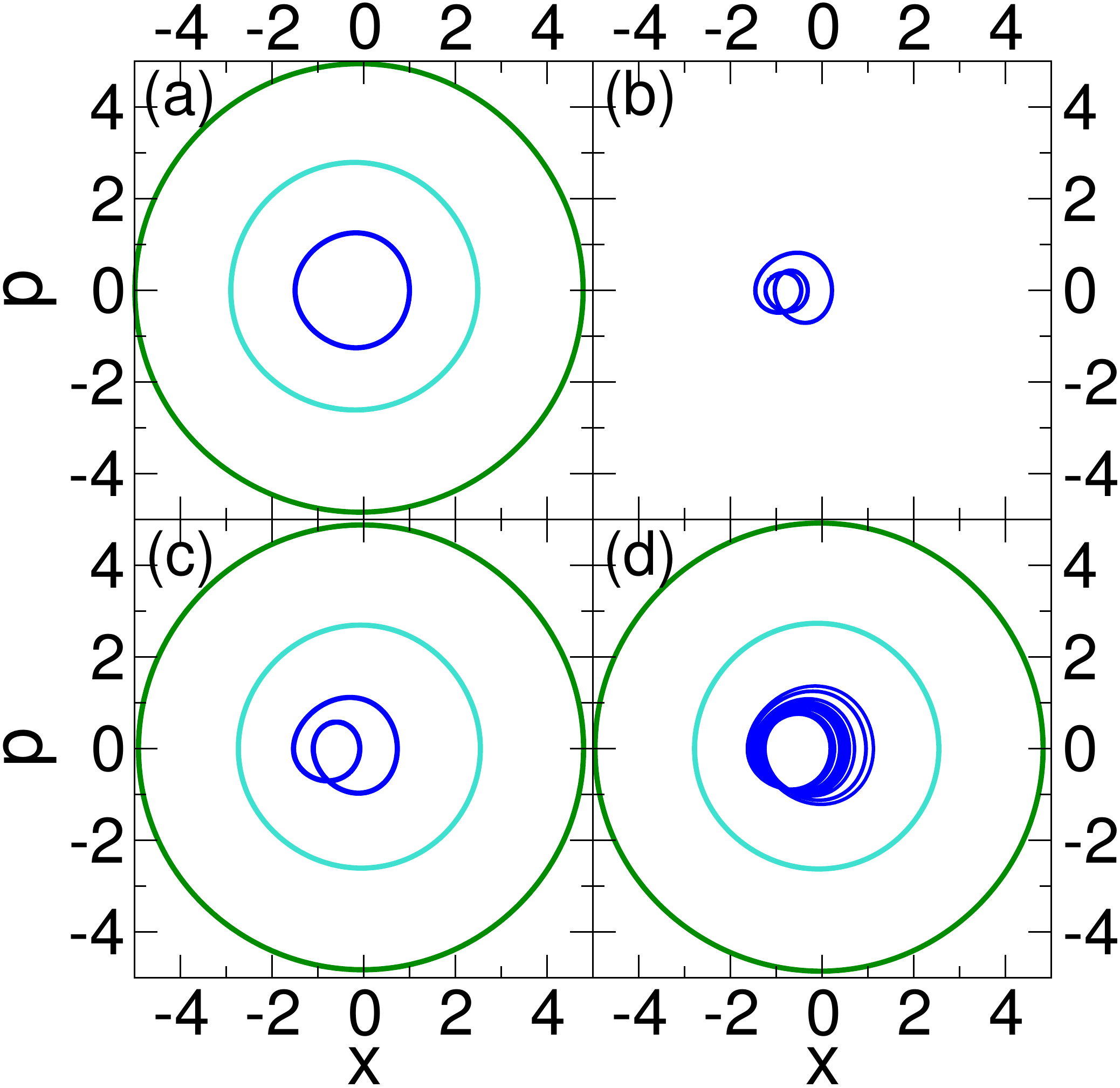}
\caption{(color online) Left panel: Chart of self-sustained oscillations in the classical limit for $P=1.5$.
Self-sustained oscillations occur for amplitudes $A$ where the power balance
between gains from the radiation pressure 
($P_\text{rad} =P \langle |\alpha|^2 \mathrm{Im} \beta  \rangle_\mathrm{avg}$)
and losses due to friction 
($P_\text{fric} =\bar\Gamma \langle |\beta|^2 \rangle_\mathrm{avg} $)
changes from positive to negative values with increasing $A$~\cite{MHG06,LKM08}.
Right panels: Classical orbits in the $(x,p)$ cantilever phase space,
for (a) $\Delta = -0.4$, (b) $\Delta = - 1.1$, (c) $\Delta = - 0.85$, and (d) $\Delta =-0.7$, as marked by vertical lines in the left panel. In case (a), the two innermost orbits have amplitudes $A_1 \approx 1.2$ and $A_2 \approx 2.7$.
In cases (b), (c) the innermost orbit shows a few period doubling bifurcations that occur on the route to chaos~\cite{BAF14_PRL}, in case (d) it is chaotic.
}
\label{fig:chart}
\end{figure}

Now introduce the five dimensionless parameters~\cite{MHG06, LKM08}
\begin{equation}\label{eq:NewParams}
\Delta = \dfrac{\Omega_\text{las} - \Omega_\text{cav}}{\Omega}   \;, \quad
 P = \dfrac{8 \alpha_\text{las}^2 g_\text{rad}^2}{\Omega^4} \;,
\quad
\sigma = \dfrac{g_\text{rad}}{\kappa} \;,
\end{equation}
and $\bar\kappa = \kappa / (2 \Omega)$, $\bar\Gamma = \Gamma / (2 \Omega)$,
and measure time as $\tau = \Omega t$.
The parameter $\Delta$ gives the detuning of the pump laser and cavity, while $P$ gives the strength of the laser pumping.
For later numerical results we set the damping parameters $\bar\kappa = 0.5$, $\bar \Gamma = 5 \times 10^{-4}$
to typical experimental values~\cite{AKM13}.

The quantum-classical scaling parameter $\sigma$ is the ratio of the quantum mechanical quantity $g_\text{rad}$, which is of order $\hbar^{1/2}$ because the quantum mechanical position operator $\hat x \propto \hbar^{1/2} (b^\dagger + b)$ of the cantilever enters the expression for the radiation pressure, to the classical quantity $\kappa$ that measures the cavity quality.
The parameter $\sigma$ thus controls the crossover from classical ($\sigma = 0$) to quantum ($\sigma > 0$) mechanics~\cite{LKM08}.
In the following we will increase $\sigma$ to move into the quantum regime,
but keep $\sigma \ll 1$ in order to remain in the vicinity of the classical limit $\sigma = 0$.

\section{Classical multistability}

Our analysis begins in the limit $\sigma = 0$,
where the optomechanical system is described by the classical equations of motion~\cite{LKM08}
\begin{subequations}
\begin{align}
 \partial_\tau \alpha & = (\ii \Delta - \bar\kappa) \alpha  - \ii (\beta + \beta^*) \alpha - \tfrac12 \ii  \;, \\[0.25ex]
 \partial_\tau \beta & = (-\ii  -\bar{\Gamma} ) \beta   - \tfrac12 \ii P |\alpha|^2  
\end{align}
\label{eq:SC}%
\end{subequations}
for the cavity and cantilever phase space variables
$\alpha = (\Omega/(2 \alpha_\text{las})) \langle a \rangle$, $\beta = (g_\text{rad}/\Omega) \langle b \rangle$.
We also use the 
cantilever position and momentum operator
$\hat x =(1/\sqrt2) (g_\text{rad}/\Omega) (b^\dagger + b)$,
$\hat p = (\ii/\sqrt2) (g_\text{rad}/\Omega) (b^\dagger - b)$,
with corresponding phase space variables $x =\langle \hat x \rangle = 1/\sqrt{2}\, (\beta + \beta^*)$ and $p =\langle \hat p \rangle = (\ii/\sqrt{2}) (\beta^* - \beta)$.

The classical equations of motion predict the onset of self-sustained cantilever oscillations 
$x(\tau) = x_0 + A \cos \tau$
as the pump power $P$ is increased. Figure~\ref{fig:chart} shows the possible amplitudes $A$ of these oscillations, which are obtained with the ansatz from Ref.~\cite{MHG06}, for the value $P=1.5$.
We keep this value fixed throughout the paper, as the behaviour discussed here does not depend on it.
Note in Fig.~\ref{fig:chart} that several stable oscillatory solutions at different amplitudes $A$ can coexist for one parameter choice.
This classical multistability of self-sustained oscillations is the origin of the quantum multistability analyzed next.

\section{Quantum multistability}

\begin{figure}
\centering
\includegraphics[width=0.55\linewidth]{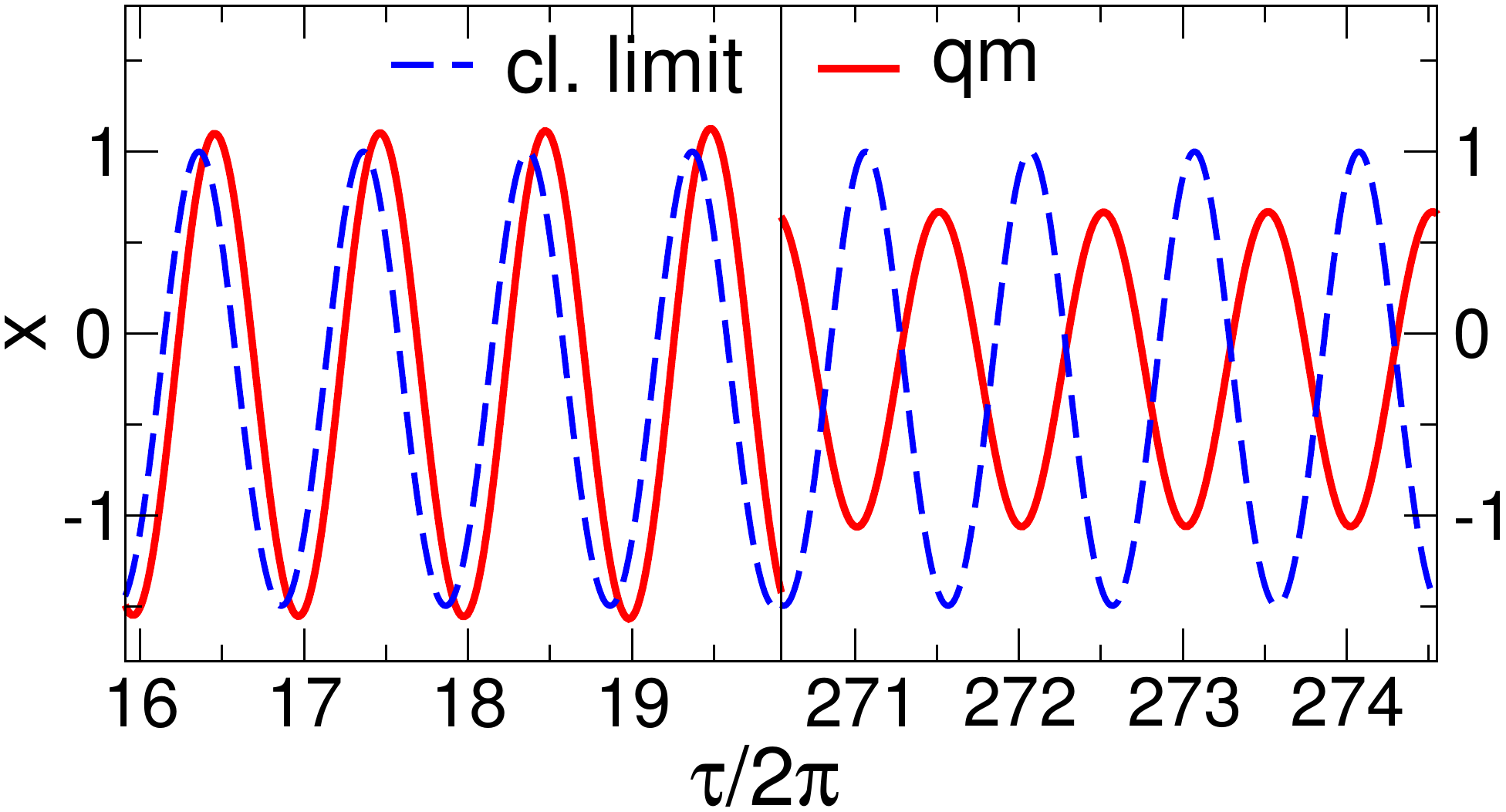}
\hfill
\includegraphics[width=0.43\linewidth]{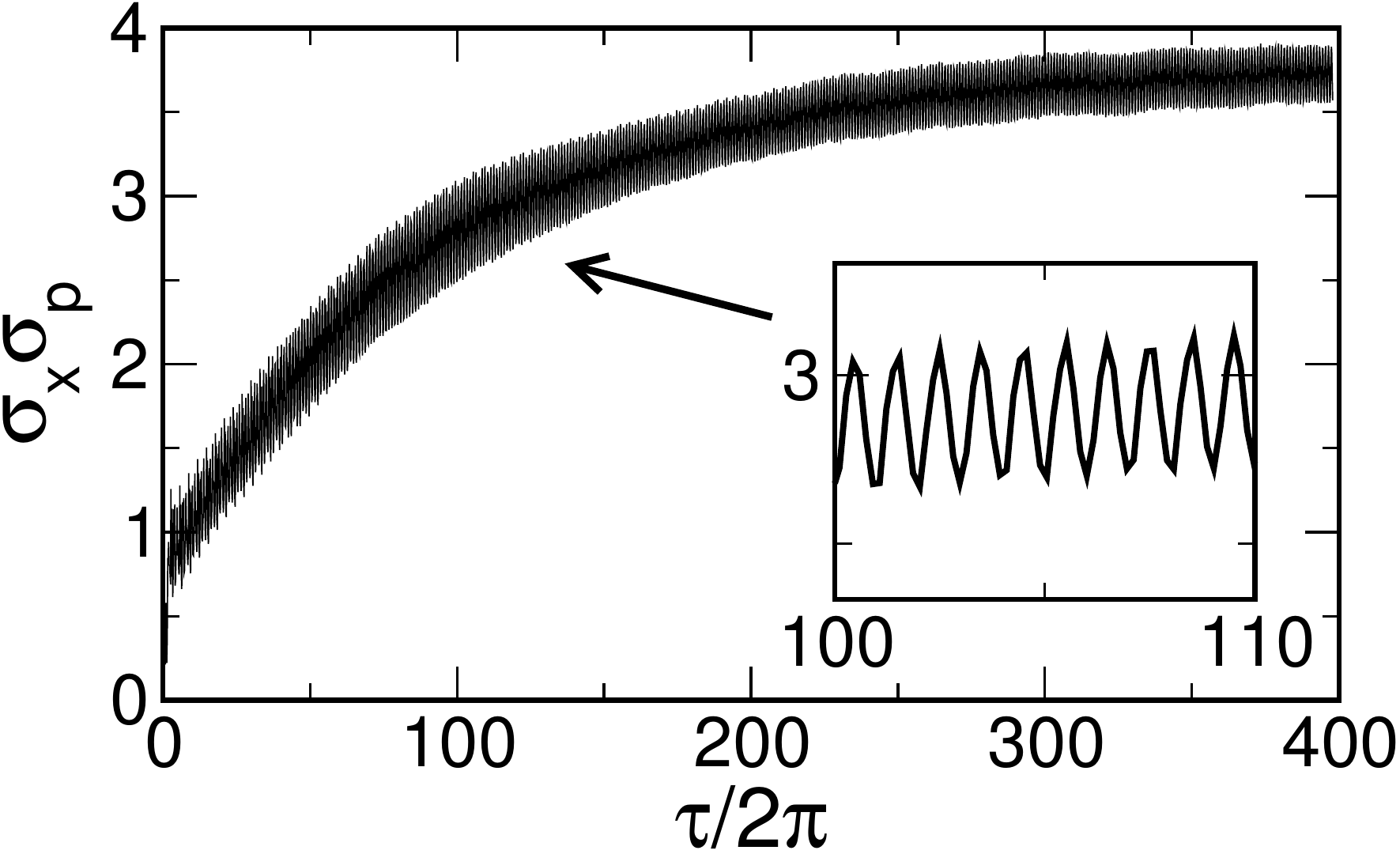}
\caption{(color online) Left panel: Cantilever position $x(\tau)$ from the classical equations of motion~\eqref{eq:SC} and from the quantum mechanical master equation~\eqref{eq:master} at $\sigma = 0.1$,
for $P=1.5$, $\Delta = -0.4$ (case (a) in Fig.~\ref{fig:chart}).
Right panel: Cantilever position-momentum uncertainty product $\sigma_x \sigma_p$ for the same parameters.
}
\label{fig:boring}
\end{figure}

We now move into the quantum regime by letting $\sigma$ become finite.
In all our examples the quantum system is initially prepared in the pure product state 
of a coherent cantilever and cavity state at $\alpha = \beta =0$, i.e., in the state that is closest to a classical state at these coordinates.
The cantilever-cavity density matrix is then evolved according to Eq.~\eqref{eq:master}.

Figure~\ref{fig:boring} shows the cantilever position $x$ and the position-momentum uncertainty product $\sigma_x \sigma_p$, with the uncertainty 
$\sigma_O = (\langle {\hat O}^2 \rangle - \langle \hat O \rangle^2)^{1/2}$
of an observable $O$.
The quantum dynamics at finite $\sigma$ closely follows the classical oscillations for an initial period of time, before it deviates significantly at later times.
Deviations occur because the quantum state spreads out in phase space,
as witnessed by the growth of the uncertainty product, whereby the cantilever position is smeared out.

\begin{figure}
\centering
\begin{minipage}{0.48\linewidth}
\centering
\includegraphics[width=0.98\linewidth]{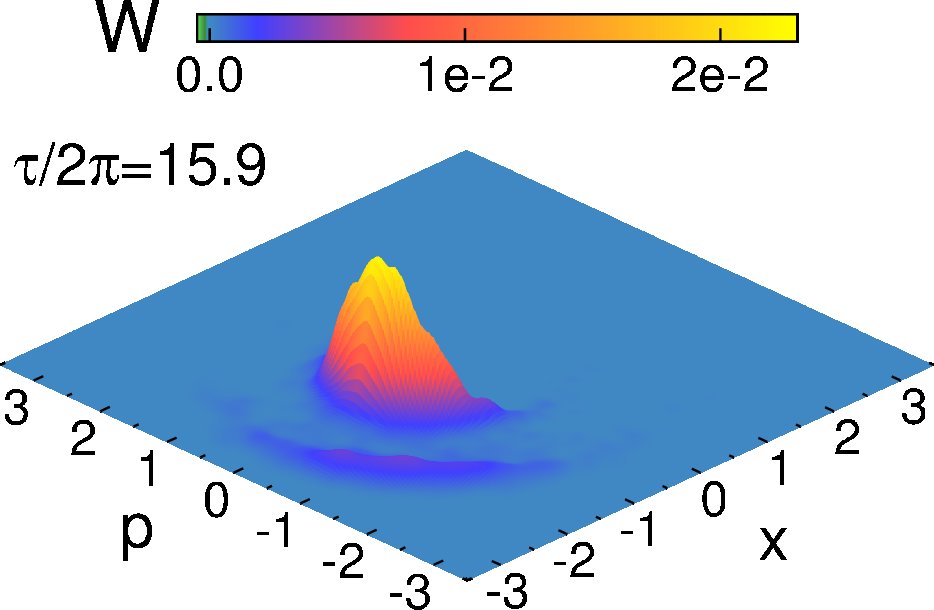} \\[1ex]
\includegraphics[width=0.98\linewidth]{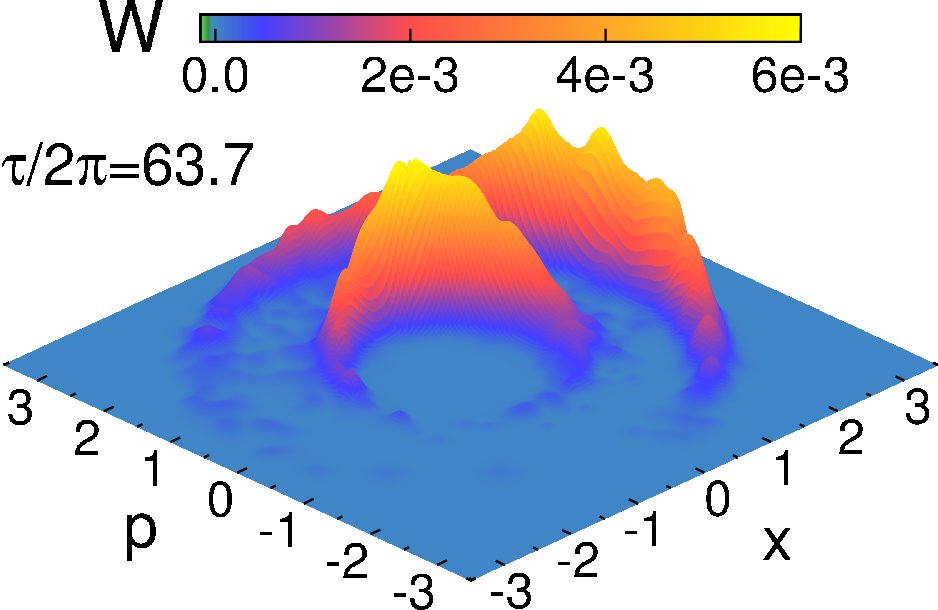} \\[1ex]
\includegraphics[width=0.98\linewidth]{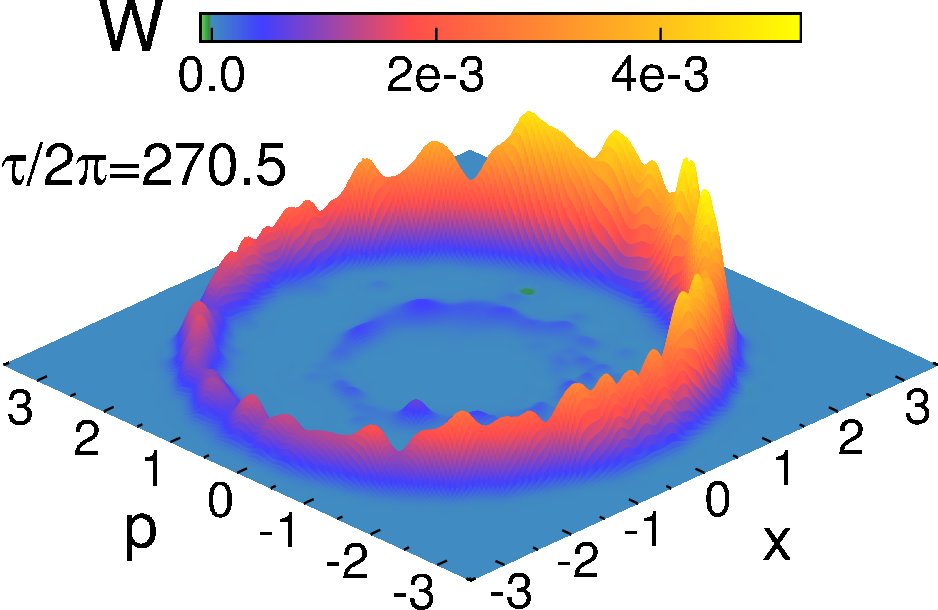} 
\end{minipage}
\begin{minipage}{0.48\linewidth}
\centering
\includegraphics[width=0.98\linewidth]{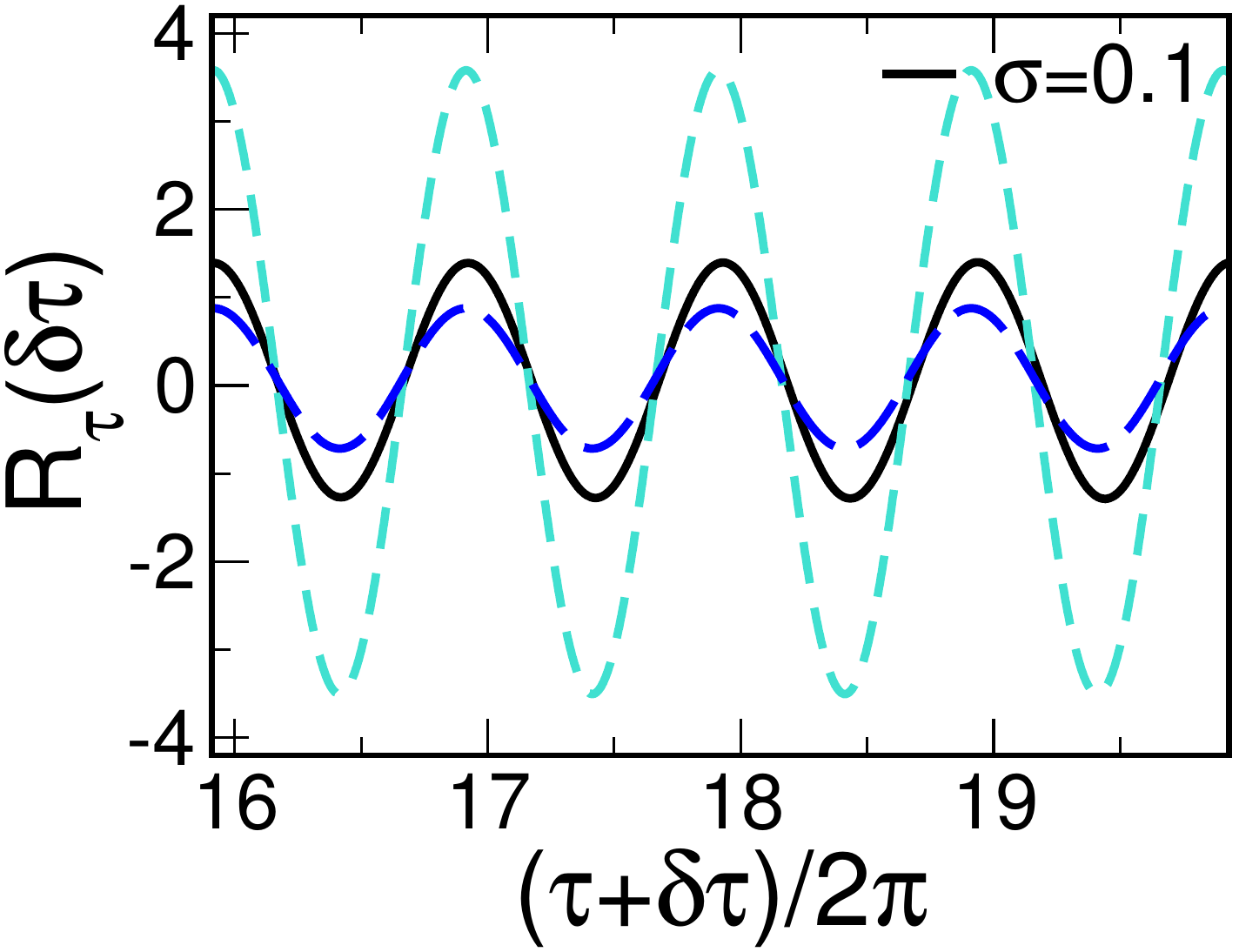} \\[1ex]
\includegraphics[width=0.98\linewidth]{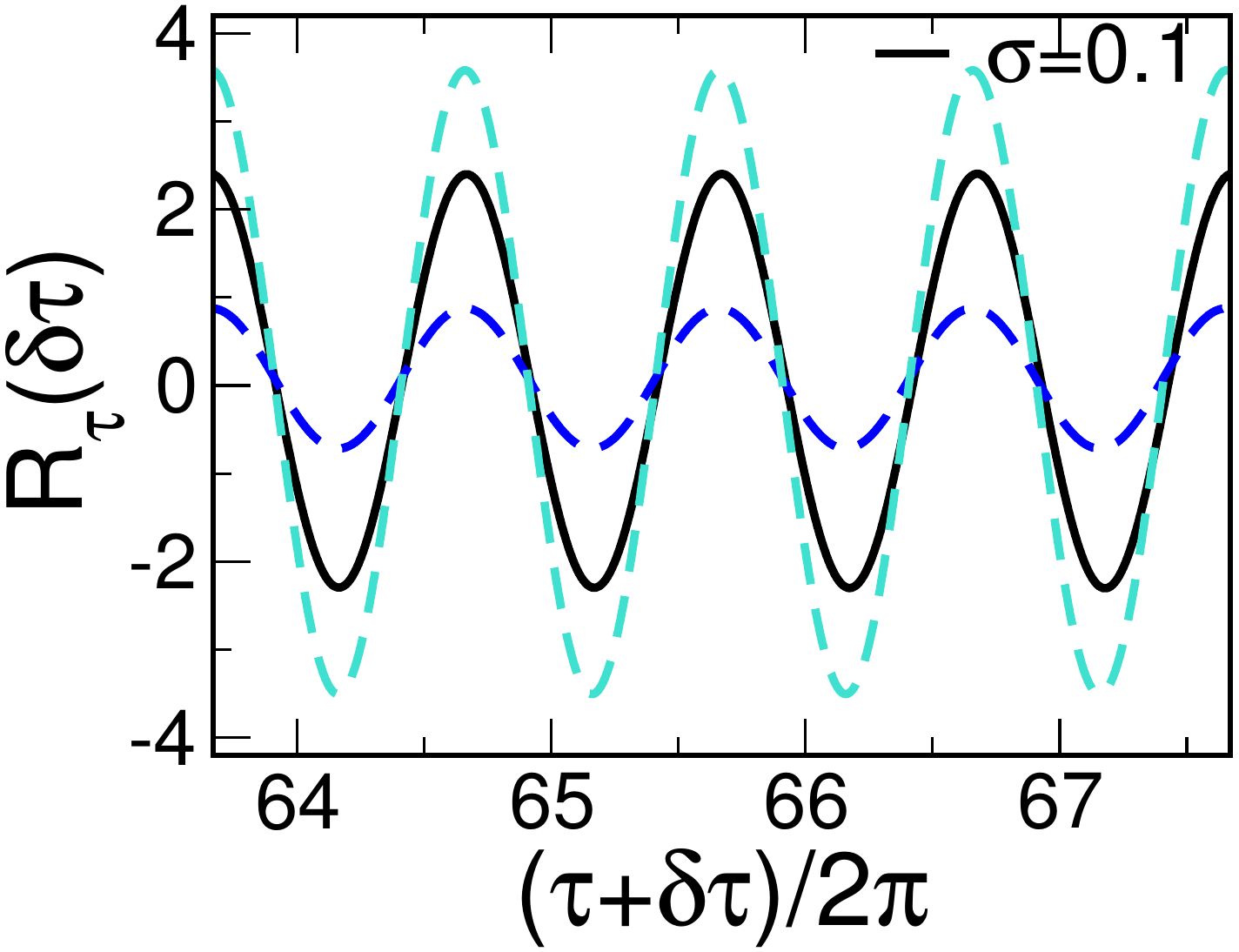} \\[1ex]
\includegraphics[width=0.98\linewidth]{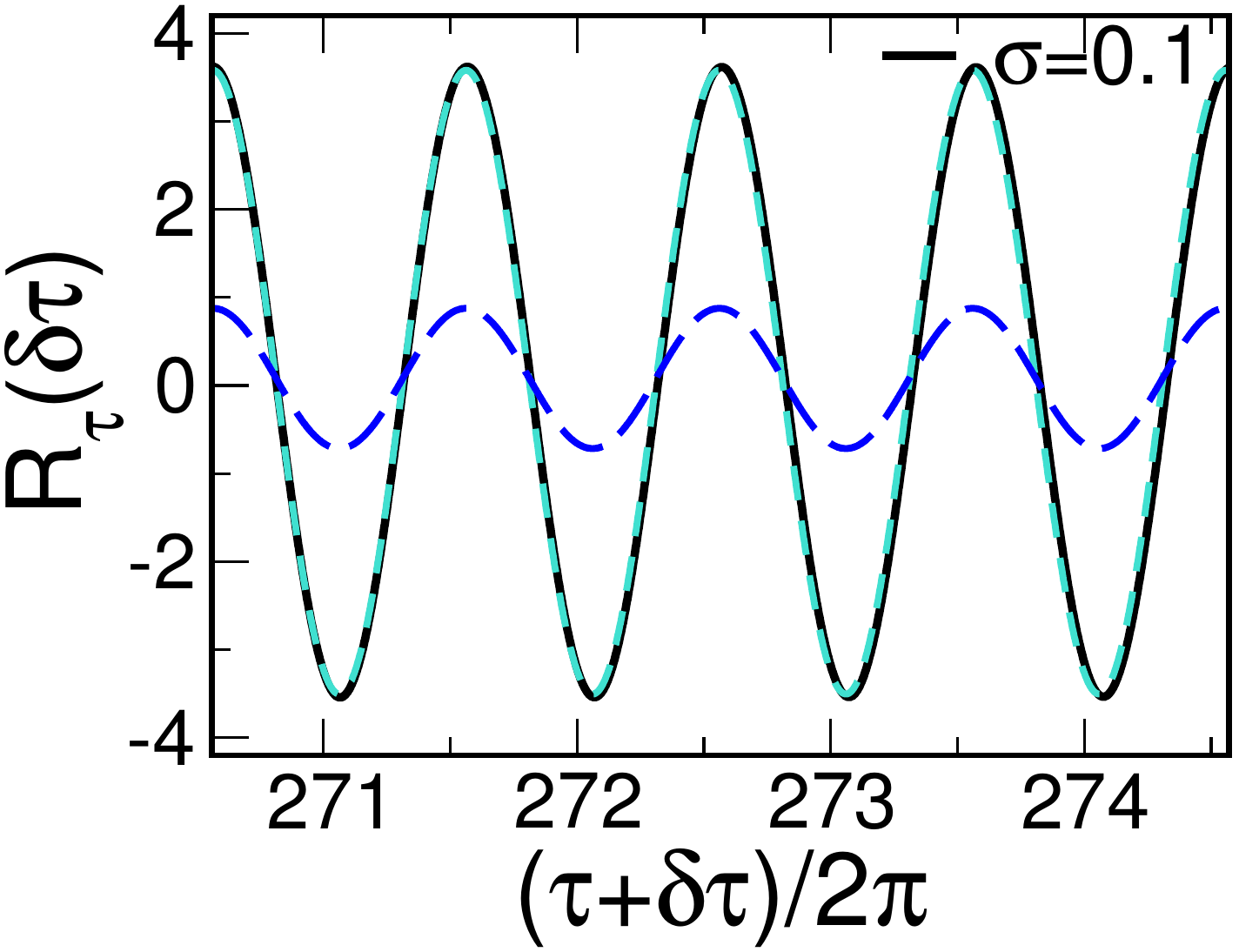}
\end{minipage}
\caption{(Color online)
Wigner function $W(x,p)$ in cantilever phase space (left panels)
and cantilever position autocorrelation function $R_\tau(\delta\tau)$ (right panels)
for case (a) from Fig.~\ref{fig:chart},
at $\sigma=0.1$ slightly away from the classical limit.
The autocorrelation functions for the two inner classical orbits at amplitudes $A_{1/2}$ are included as dashed curves.}
\label{fig:Wigner}
\end{figure}

The full phase space dynamics in Fig.~\ref{fig:Wigner},
which we display with the Wigner function $W(x,p)$ of the cantilever mode (see, e.g., Ref.~\cite{Schl01} for the definitions), reveals a more definite dynamical pattern.
For early times ($\tau \simeq 16$) the Wigner function retraces the classical orbit with amplitude $A_1 \approx 1.2$ from case (a) in Fig.~\ref{fig:chart}.
At later times ($\tau \simeq 64$) the Wigner function shows a contribution from a second circular orbit with larger amplitude,
before almost all weight is concentrated on the new orbit ($\tau \simeq 270$).
In comparison to case (a) in Fig.~\ref{fig:chart} this orbit is identified as the second classical orbit with amplitude $A_2 \approx 2.7$.
During time evolution the quantum state spreads out along, but not perpendicular to, these two classical orbits.

The classical multistability of the optomechanical system thus has a direct counterpart in the quantum dynamics at small $\sigma$,
where the system moves between the different classical orbits.
This kind of quantum multistability leads to distinct dynamical features because the oscillatory nature of the different orbits is preserved.

The quantum multistability is clearly detected with the cantilever position autocorrelation function
\begin{equation}\label{eq:auto}
R_\tau(\delta \tau) = \int\limits_{\tau-\pi}^{\tau+\pi} \langle \hat x(\tau') \hat x(\tau'+\delta \tau) \rangle \, d\tau'  \;,
\end{equation}
instead of the position expectation value that averages over the phase space distribution.
We choose this function because the dynamics is best described in cantilever phase space. Autocorrelation functions for the cavity mode could be used as well
and should be more accessible to experimental measurements,
but their interpretation is less straightforward because of the additional sidebands at multiples of the fundamental oscillation frequency.

The autocorrelation function in Fig.~\ref{fig:Wigner} is the weighted sum of the oscillatory motion on the two orbits seen in the Wigner function.
The frequency of the two orbits is identical (essentially, the cantilever frequency $\Omega$), such that only one oscillation is visible in $R_\tau(\delta \tau)$.
The amplitude of $R_\tau(\delta \tau)$ increases as weight is transferred from the inner to the outer orbit.
Noteworthy, the oscillations persist at all times.
In this way, the multistability of the quantum dynamics 
is not only observable during a short initial time period 
but during extended periods of time.

\section{Multistability of quantum trajectories}

The mechanism behind the quantum multistability can be understood through the phase space dynamics of individual quantum trajectories, as they arise in the quantum state diffusion (QSD) approach~\cite{GP92} to the solution of Lindblad master equations such as Eq.~\eqref{eq:master}.

\begin{figure}
\centering
\includegraphics[width=0.48\linewidth]{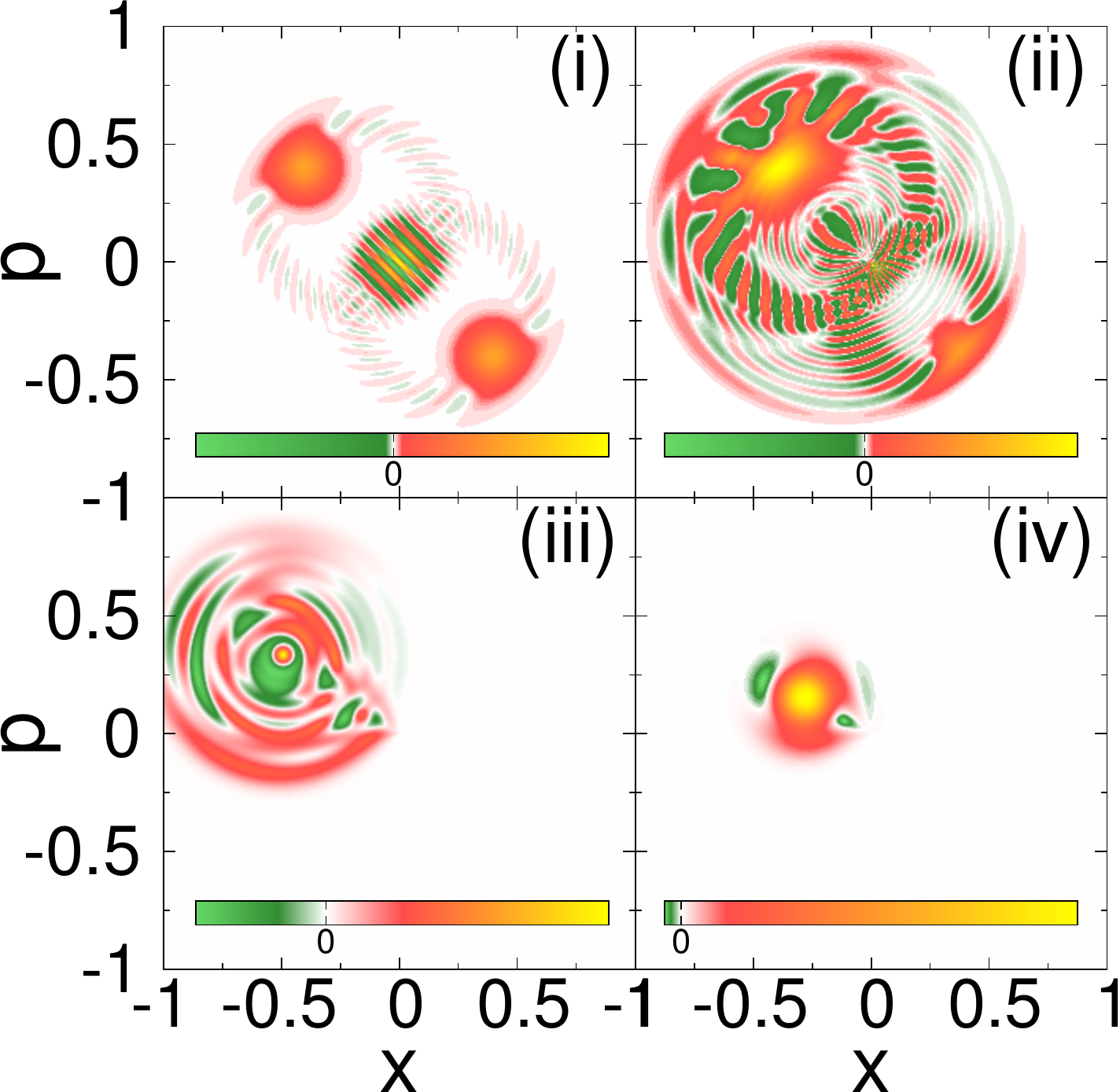}
\includegraphics[width=0.48\linewidth]{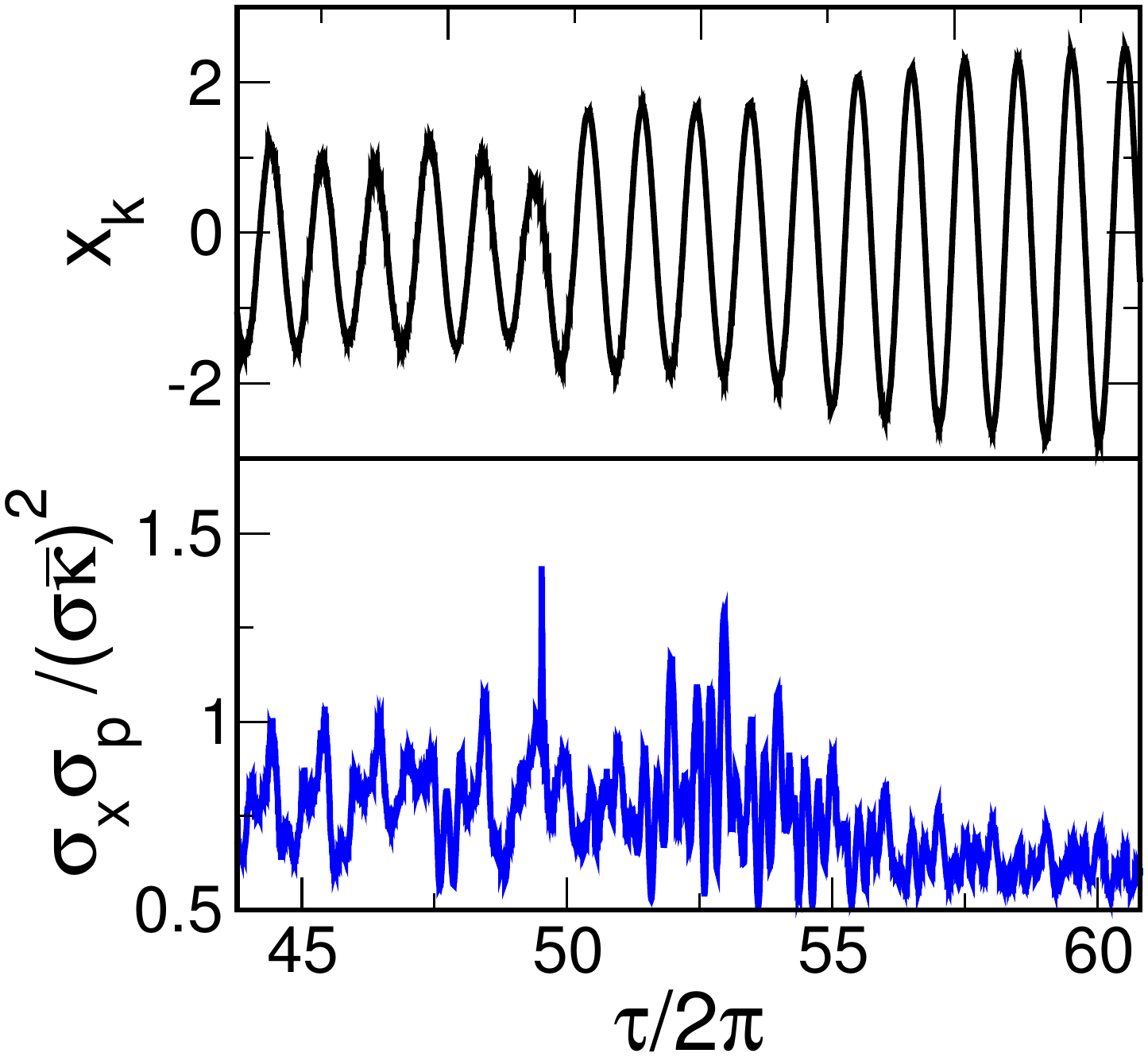}
\caption{(color online) 
Left panels: Wigner function $W(x,p)$ for a single quantum trajectory starting from a  ``Schr\"odinger cat'' state at (i) $\tau = 0$,
and at later times (ii) $\tau =0.001$, (iii) $\tau = 0.008$, and (iv) $\tau = 0.4$.
Right panels:
Cantilever position $x_k$ and uncertainty product $\sigma_x \sigma_p$ (see Eq.~\eqref{eq:Heisenberg}) for a single quantum trajectory at later times, in the situation of Fig.~\ref{fig:Trajectory}.
All results are for case (a) from Fig.~\ref{fig:chart} and $\sigma = 0.1$.
}
\label{fig:uncertain}
\end{figure}

\begin{figure}
\centering
\includegraphics[width=0.8\linewidth]{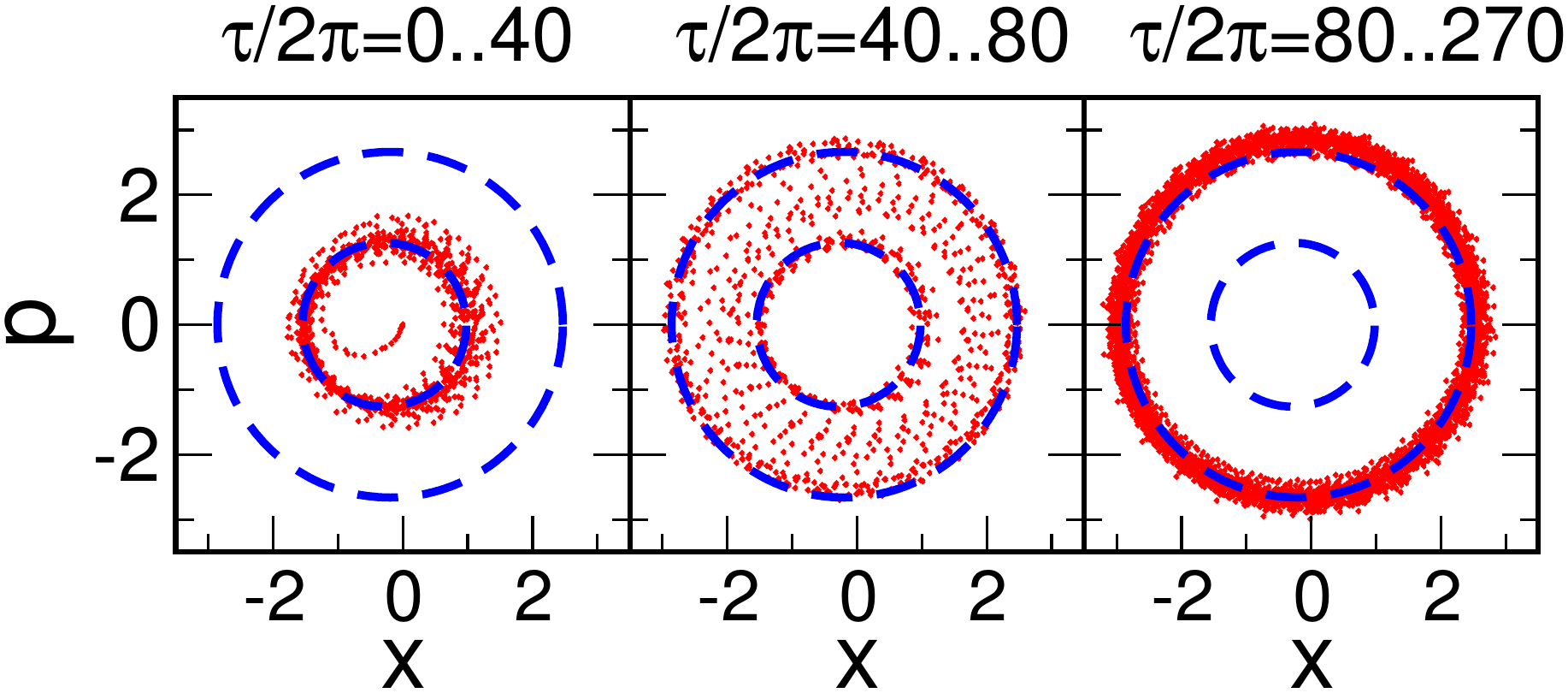}
\caption{(color online) Stroboscopic $(x,p)$-phase space plot of a single quantum trajectory (red dots), for case (a) from Fig.~\ref{fig:chart} and $\sigma=0.1$, at early (left panel), intermediate (central panel), and later (right panel) times $\tau$ as indicated.
The initial conditions are $x(0)=p(0)=0$, the quantum system is prepared in a coherent state at these coordinates.
The two classical orbits at amplitudes $A_{1/2}$ are depicted with dashed curves.
}
\label{fig:Trajectory}
\end{figure}

\begin{figure}
\centering
\includegraphics[width=0.92\linewidth]{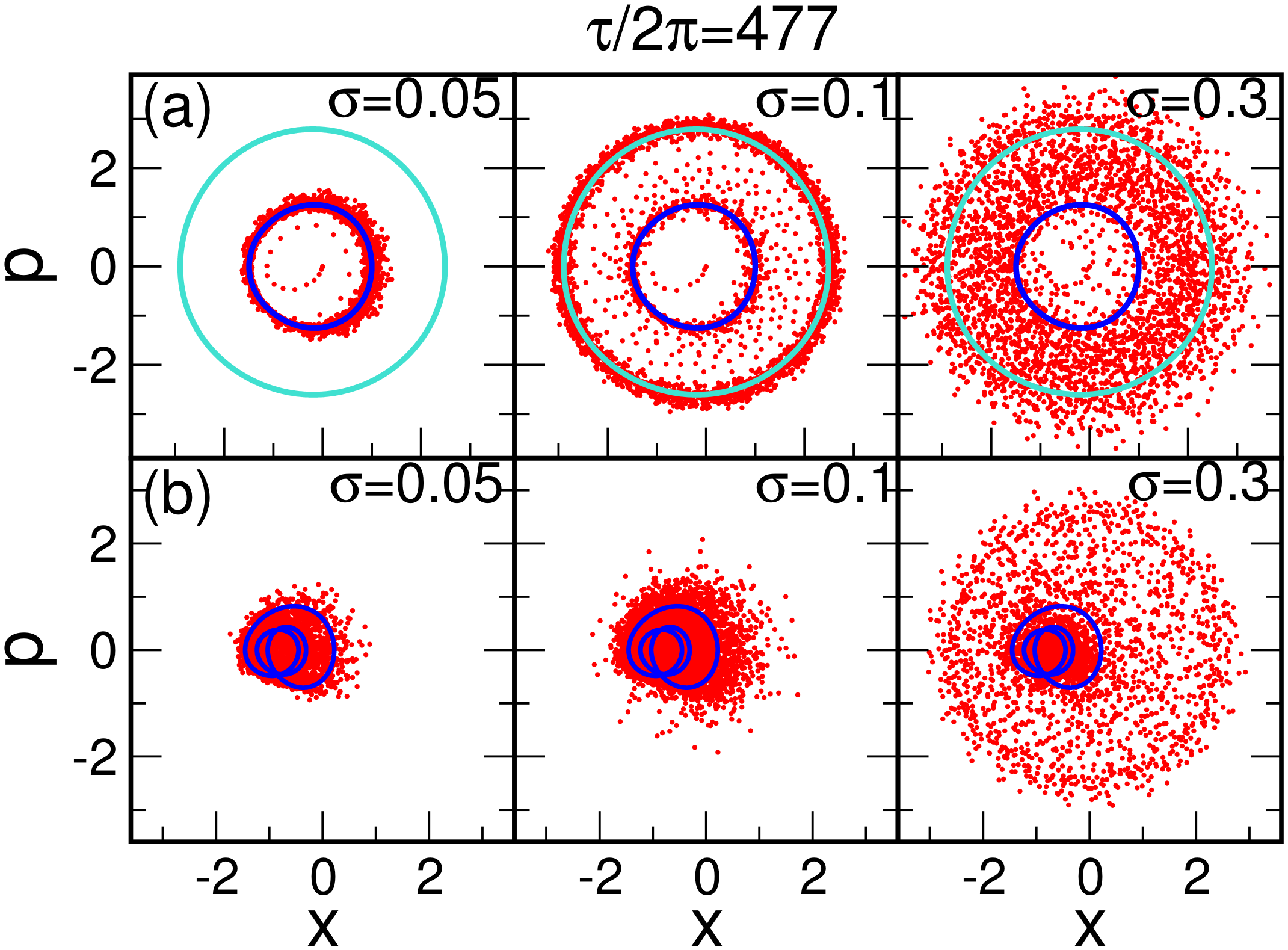} \\
\includegraphics[width=0.92\linewidth]{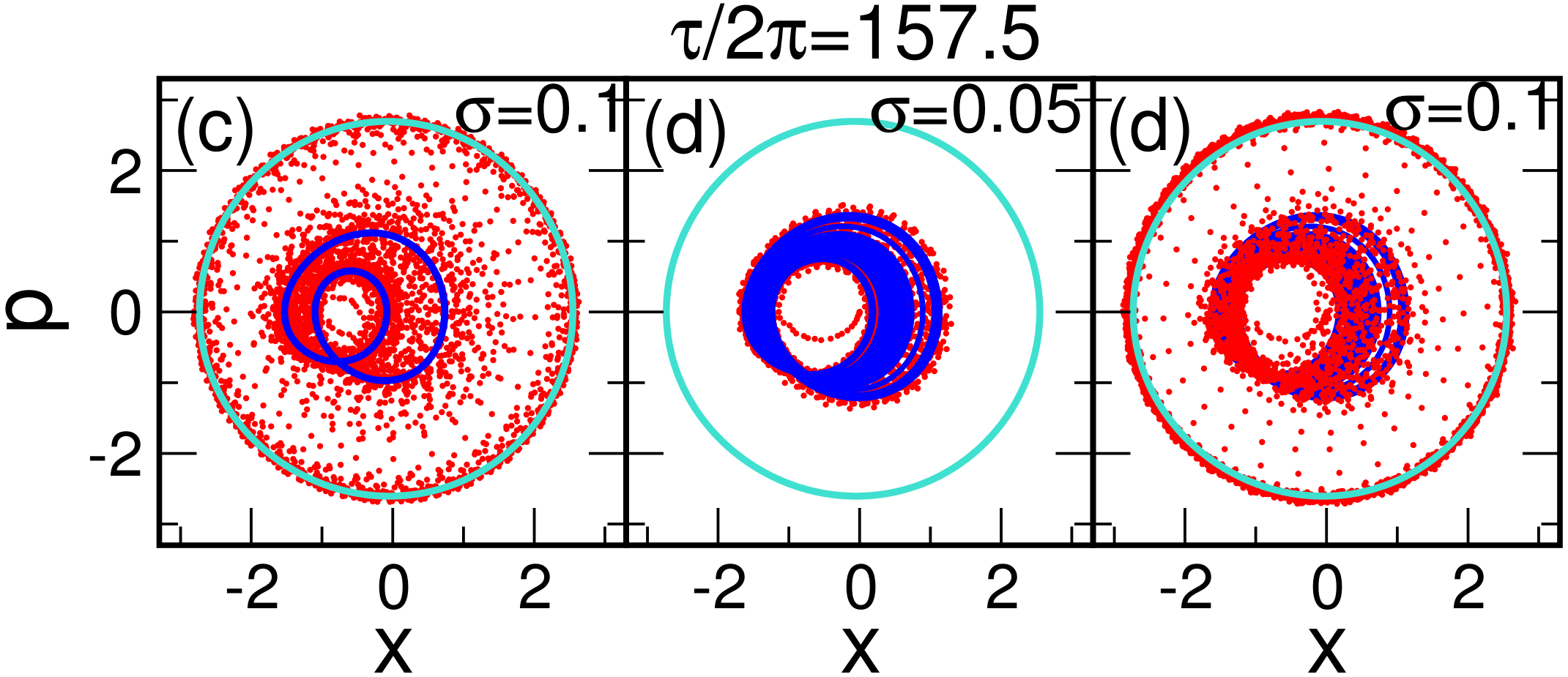}
\caption{(color online) Phase space plot of many quantum trajectories in the QSD ensemble (red points) for 
cases (a)--(d) from Fig.~\ref{fig:chart}, different times $\tau$, and values of $\sigma$ as indicated.
In all cases, the two innermost classical orbits from Fig.~\ref{fig:chart} are included as solid curves.
In case (b) the second orbit is missing.
}
\label{fig:TrajectoriesOther}
\end{figure}

In QSD the density matrix is represented by an ensemble of quantum trajectories $|\psi_k(\tau) \rangle$,
from which it is obtained as an average
\begin{equation}\label{eq:mean}
 \rho(\tau) = \textsf{mean}_k \Big\{ |\psi_k(\tau) \rangle \langle \psi_k(\tau) | \Big\} \;.
\end{equation}
Accordingly, expectation values are computed as ensemble averages 
$O(\tau) = \tr [ \rho(\tau) \hat O ] =  \textsf{mean}_k \Big\{  \langle \psi_k(\tau) | \hat O | \psi_k(\tau) \rangle \Big\} $.
Each quantum trajectory $|\psi_k(\tau) \rangle$ follows a stochastic equation of motion that combines the Hamiltonian and dissipative dynamics with a noise term~\cite{GP92}.
Numerically, the density matrix is obtained through Monte Carlo sampling of the trajectories for different noise realizations. We use the QSD implementation from Ref.~\cite{SB97},
and typically average over $\simeq 3000$ trajectories to obtain the results in 
Figs.~\ref{fig:boring}--\ref{fig:uncertain},~\ref{fig:WignerOthers},~\ref{fig:AutoOthers}.
Although a single quantum trajectory is not observable by itself,
the phase space dynamics of individual trajectories as shown in Figs.~\ref{fig:Trajectory},~\ref{fig:TrajectoriesOther} allows us to deduce the properties of the entire density matrix.

Close to the classical limit quantum trajectories evolve rapidly into localized phase space states as a consequence of dissipation~\cite{Per94,SBP95,RG96}.
This is illustrated in Fig.~\ref{fig:uncertain} for a single trajectory that starts from a ``Schr\"odinger cat'' state, given as the superposition of two coherent states, with the characteristic interference pattern in the Wigner function.
In less than one oscillation period ($\tau =0.4 < 1$) the trajectory evolves into a nearly coherent state with a positive Wigner function, which shows the rapid decoherence.
The quantum trajectory remains in such a state during the subsequent  time evolution,
and the uncertainty product stays close to its minimal value
\begin{equation}\label{eq:Heisenberg}
 \sigma_q \sigma_p \ge \tfrac12 (g_\text{rad}/\Omega)^2 = \tfrac12 (\sigma \bar\kappa)^2
\end{equation}
given by the Heisenberg uncertainty relation for the $\hat x$, $\hat p$ operators (here, the quantum-classical scaling parameter $\sigma$ comes into play).
Notice that phase space localization occurs only in the vicinity of the classical limit, for $\sigma \ll 1$.
It also explains the transition into the classical limit:
For $\sigma \to 0$ the quantum trajectories evolve infinitely fast into minimal uncertainty states,
and at the same time the lower bound in Eq.~\eqref{eq:Heisenberg} goes to zero.
Then, every trajectory occupies one point in phase space, i.e., it has become classical.
Under this condition the classical equations of motion~\eqref{eq:SC} 
can be derived directly from the master equation~\eqref{eq:master}.

Because a quantum trajectory is very localized in phase space it is well represented by a single phase space point, similar to a classical trajectory.
In Fig.~\ref{fig:Trajectory} this representation is used for a stroboscopic phase space plot of a single quantum trajectory that contributes to the Wigner functions in Fig.~\ref{fig:Wigner}.
This plot clearly shows the multistability of the quantum trajectory,
which initially follows the inner orbit before it moves towards the outer orbit.
During the time evolution the quantum trajectory follows the oscillatory motion of the two orbits at the cantilever frequency, and because the trajectory state is well localized in phase space these oscillations are not averaged out but appear directly in the position expectation value $x_k(\tau) = \langle \psi_k(\tau) | \hat x  | \psi_k(\tau) \rangle$ that is depicted in Fig.~\ref{fig:uncertain}.

Since every individual trajectory shows this type of quantum multistability it is also seen in the entire density matrix,
given as the ensemble average of all trajectories.
Because of the noise term in the stochastic QSD equation of motion the quantum trajectories are not exactly at the same phase space point but at different points on the respective orbits. This results in the broad distribution of the relative angle in phase space seen in the Wigner functions in Fig.~\ref{fig:Wigner} especially at later times,
when the quantum trajectories are spread out fully along the second orbit.
Consequently, all oscillations are averaged out in expectation values such as the cantilever position $x(\tau)$ in Fig.~\ref{fig:boring}. Such values are, therefore, not the right quantities to detect the quantum multistability.

Instead, successful detection requires autocorrelation functions such as $R_\tau(\delta\tau)$ from Eq.~\eqref{eq:auto}.
Similar to the density matrix the function $R_\tau(\delta\tau)$ can be expressed (dropping the $\tau'$-integration here) as an ensemble average
\begin{multline}\label{eq:autocorr}
R_\tau(\delta\tau) =  \sum_k x_k(\tau) \, x_k(\tau+\delta\tau) \\
 + \sum_k \big\langle \big(\hat x(\tau) - x_k(\tau) \big) \big( \hat x(\tau+\delta\tau) - x_k(\tau+\delta\tau) \big)  \big\rangle_k \;,
\end{multline}
where the expectation value $\langle \cdot \rangle_k = \langle \psi_k | \cdot | \psi_k \rangle$ is computed for each individual quantum trajectory.
The correlation function in the second line is bounded by
\begin{multline}
 \Big| \big\langle \big(\hat x(\tau) - x_k(\tau) \big) \big( \hat x(\tau+\delta\tau) - x_k(\tau+\delta\tau) \big)  \big\rangle_k \Big|^2  \\
 \le \big\langle \big(\hat x(\tau) - x_k(\tau) \big)^2 \big\rangle_k \,
\big\langle \big(\hat x(\tau+\delta\tau) - x_k(\tau+\delta\tau) \big)^2 \big\rangle_k \;.
\end{multline}
Whenever the position uncertainty $\langle (\hat x - x_k)^2 \rangle_k$ of each trajectory becomes small,
as it is the case for $\sigma \ll 1$, the autocorrelation function $R_\tau(\delta\tau)$  is thus given by the ensemble average of the autocorrelation functions of the individual trajectories, i.e., by the first line in Eq.~\eqref{eq:autocorr}.
Accordingly, the oscillations seen in $x_k(\tau)$ for each individual trajectory (cf. Fig.~\ref{fig:uncertain}) are preserved in the autocorrelation function in spite of the ensemble average.
Furthermore, $R_\tau(\delta\tau)$ is the weighted sum of the autocorrelation functions for the different classical orbits, which are directly related to the orbit amplitudes $A_{1/2}$ as seen in Fig.~\ref{fig:Wigner}.

Notice that the behaviour described here---the motion of quantum trajectories between different classical orbits---emerges only because the trajectory states $|\psi_k\rangle$ deviate from coherent states.
The noise terms in the QSD equations have the form $\bar\Gamma (b - \langle b \rangle_k) |\psi_k\rangle \mathrm d \xi$, here for the mechanical damping, with a random variable $\mathrm d\xi \propto \mathrm d\tau^{1/2}$ from the underlying Wiener process~\cite{GP92}.
If $|\psi_k\rangle$ is exactly a coherent state, such that $(b - \langle b \rangle_k) |\psi_k\rangle = 0$,
the noise term will vanish identically.
This observation explains why the ``quantum noise'' disappears in the classical limit $\sigma=0$,
and the quantum trajectories follow the deterministic classical equations of motion~\eqref{eq:SC}.
At finite but small $\sigma \ll 1$ trajectories are almost but not exactly in coherent states.
The noise terms become effective but remain small,
such that the quantum trajectories still follow the classical dynamics but are subject to a small stochastic correction.
This small correction can change the long-time stability of classical orbits and their basin of attraction but does not destroy the classical dynamical patterns.
Consequently, the quantum trajectories do not move arbitrarily in phase space but follow a classical orbit for some time before they leave the orbit with a finite probability.
Afterwards, the trajectories can settle on a different attractive orbit if such an orbit exists at larger amplitudes.

\section{Quantum multistability and classical orbits}

The quantum multistability observed for case (a) from Fig.~\ref{fig:chart} depends on the presence of at least two classical orbits between which the quantum trajectories can move.
The remaining cases (b)--(d) are variations of this situation, where either the second orbit is missing (case (b)) or the nature of the first orbit has changed (cases (c), (d)).
The four cases are compared in Fig.~\ref{fig:TrajectoriesOther} with phase space plots of many quantum trajectories that represent the QSD ensemble for the density matrix.

In all cases the time scale relevant for quantum multistability shortens with increasing $\sigma$   because the quantum trajectories leave the first classical orbit more rapidly when the noise terms become larger.
For too large $\sigma$ (e.g., $\sigma = 0.3$ in case (a)) the clear dynamical pattern of quantum multistability---the movement between different classical orbits---disappears altogether.

In case (b) the quantum trajectories cannot settle on a nearby classical orbit once they left the first orbit. Quantum multistability, which is characterized by the prevalence of oscillatory motion over random diffusion, cannot be observed in such a situation.

\begin{figure}
\centering
\includegraphics[width=0.47\linewidth]{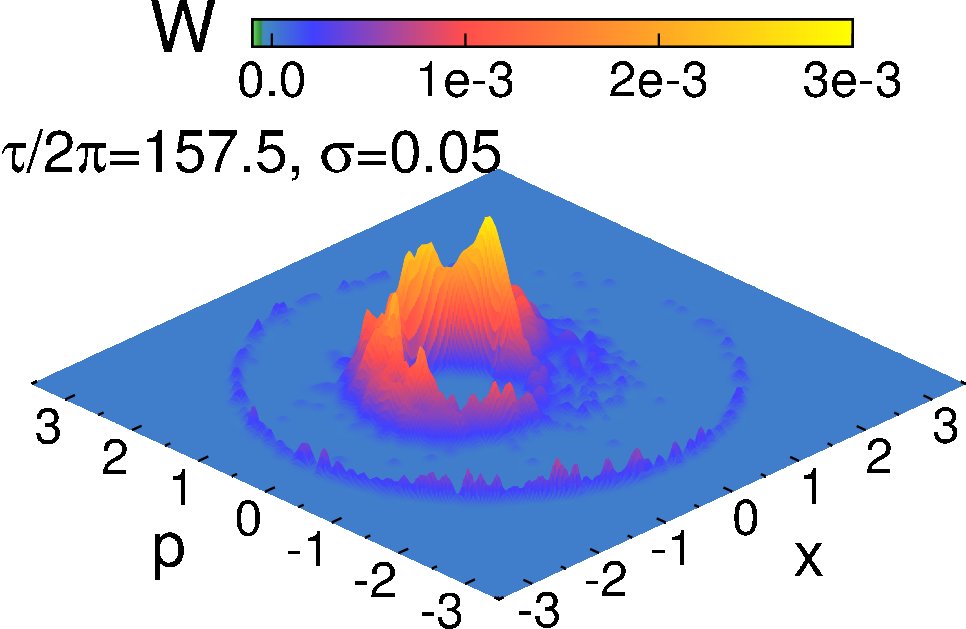}
\includegraphics[width=0.47\linewidth]{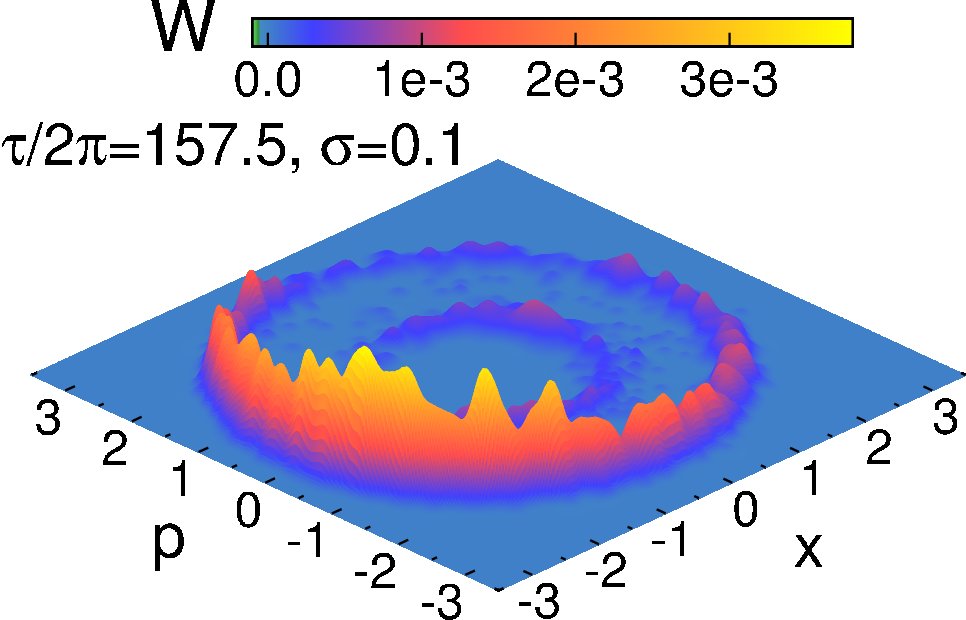}
\caption{(color online) Wigner function $W\left(x,p\right)$ in cantilever phase space for case (d) from Fig.~\ref{fig:chart}, 
for $\tau$ and $\sigma$ as in Fig.~\ref{fig:TrajectoriesOther}. 
}
\label{fig:WignerOthers}
\end{figure}

In cases (c), (d) the inner orbit is no longer simpler periodic but a period-two orbit after the first period doubling bifurcation on the route to chaos (case (c)) or a chaotic orbit (case (d)).
Quantum multistability is not affected by the different nature of the inner classical orbit,
because still a second simple periodic orbits at larger amplitude exists
such that oscillations can be observed after the quantum trajectories have left the inner orbit.

\begin{figure}
\centering
\centering
\includegraphics[width=0.92\linewidth]{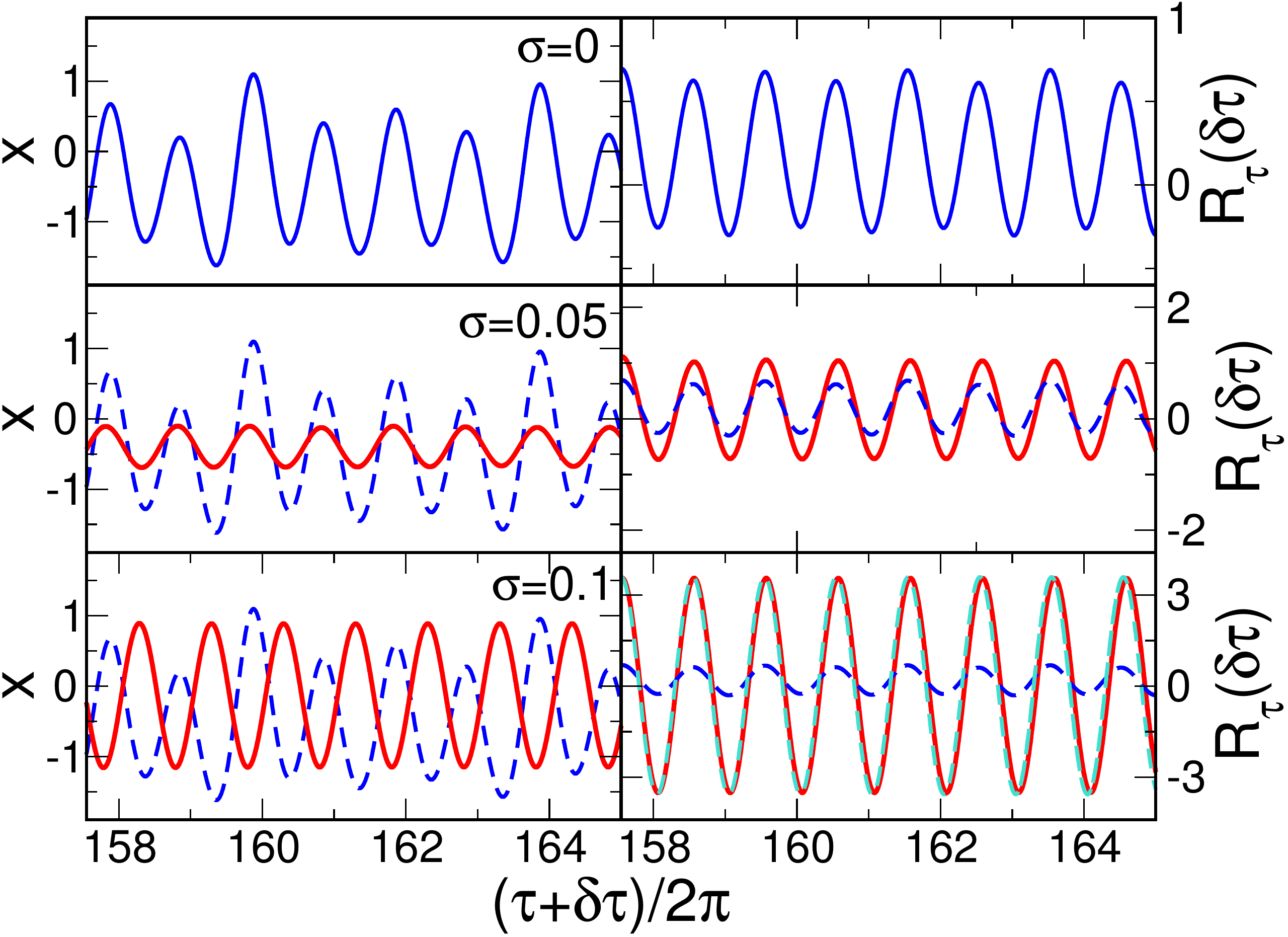}
\caption{(color online) Cantilever position $x(\tau)$ (left panels) and position autocorrelation function $R_\tau(\delta \tau)$  (right panels) for case (d) at finite $\sigma$, in comparison to the results in the classical limit $\sigma = 0$ (top panels, and dashed curves in the lower panels). These curves correspond to the Wigner functions in Fig.~\ref{fig:WignerOthers}.}
\label{fig:AutoOthers}
\end{figure}

This is illustrated for case (d) in Figs.~\ref{fig:WignerOthers},~\ref{fig:AutoOthers}.
First, we observe again that the relevant time scale changes significantly with $\sigma$.
If  $\sigma$ is increased from $0.05$ to $0.1$ in Fig.~\ref{fig:WignerOthers} almost all weight of the Wigner function is transferred from the inner to the outer orbit.
Second, the Wigner functions themselves look quite similar to those for case (a) in Fig.~\ref{fig:Wigner}.
In agreement with this, well-defined oscillations are observed in the cantilever position and autocorrelation function in Fig.~\ref{fig:AutoOthers}, and the respective amplitudes can be related to those of the classical orbits in Fig.~\ref{fig:chart}.

The present data might suggest a more ambitious interpretation.
Apparently, all curves at finite $\sigma$ in Fig.~\ref{fig:AutoOthers} show simple periodic oscillations even if (at $\sigma =0.05$) most weight in the Wigner function is still on the inner---classically chaotic---orbit.
To a certain extent, quantum mechanics protects the optomechanical system against classical chaotic dynamics.
Initially, the quantum state cannot follow the intricate chaotic orbit curve because it occupies a finite part of phase space. Because of phase space averaging the chaotic motion is replaced by simple oscillations at the fundamental system (i.e., cantilever) frequency.
Later, the quantum trajectories move to the second---classically simple periodic---orbit.
At all times, the chaotic classical dynamics is replaced by clearly defined simple oscillations in the quantum regime. Notice that we here discuss possible signatures of classical chaos in the associated dissipative quantum dynamics and not in quantities such as the level statistics that are defined for conservative Hamiltonian systems only~\cite{Gutz90,Haake10}.

\section{Conclusions}

In this paper we establish the quantum mechanical counterpart of the classical multistability of optomechanical systems. 
While classical multistability corresponds to the coexistence of self-sustained oscillations at multiple amplitudes,
quantum multistability is a dynamical effect in which the amplitude of oscillations changes over time.
The change can be detected with phase space techniques such as the Wigner function, and analyzed quantitatively with autocorrelation functions.

Quantum multistability is observed close to the classical limit.
There, the quantum trajectories in the QSD picture of dissipative dynamics are well localized in phase space.
Quantum multistability results from corrections to the classical dynamics given by the noise terms in the stochastic QSD equations of motion. 
The picture of quantum trajectories also provides the link between the oscillatory quantum dynamics and the classical orbits such that, e.g., the oscillations in the autocorrelation functions can be traced back to the classical self-sustained oscillations.

The time scale relevant for quantum multistability is set by the quantum-classical scaling parameter $\sigma$.
An interesting goal is to obtain the time scale from the QSD equations by quantifying the size of the noise term.
This is not an entirely trivial task, though, because the noise term depends not directly on $\sigma$ but on the deviation of the quantum trajectory state from a coherent state.

An important aspect for experimental investigations of quantum multistability is the robustness of the feature. Quantum multistability manifests itself over an extended period of time, is observable in autocorrelation functions after the initial dynamics has evolved into a stable dynamical pattern,
and does not require specific system preparations. The experimental feasibility depends mainly on the ability to tune the quantum-classical scaling parameter $\sigma$.
For the prototypical cantilever-cavity system $\sigma$ is changed, e.g., by simultaneous adjustment of the cantilever mass and pump laser power (thus preserving the self-sustained oscillations).
The central experimental challenge is to distinguish ``quantum'' multistability from the effects of ``classical'' thermal noise, which requires that the temperature be sufficiently low.
The relevant dynamical energies are larger than the energy separation of low-lying quantum states,
which allows for comparatively high temperatures.
Furthermore, variation of $\sigma$ changes the quantum mechanical time scale while the thermal noise is not affected.
This might open up the possibility of observing the crossover from classical to quantum mechanics directly in the dynamical behaviour of an optomechanical system.

\acknowledgments
This work was supported by Deutsche Forschungsgemeinschaft via Sonderforschungsbereich 652 (project B5).

\end{document}